\documentclass[prb,twocolumn,showpacs,preprintnumbers,amsmath,amssymb,superscriptaddress]{revtex4}

\usepackage{graphicx}
\usepackage{dcolumn}
\usepackage{color}

\newcommand{\beq}{\begin{equation}}
\newcommand{\eeq}{\end{equation}}
\newcommand{\beqa}{\begin{eqnarray}}
\newcommand{\eeqa}{\end{eqnarray}}

\begin{document}
\title{
Static vs.\ dynamical mean field theory of Mott antiferromagnets
}

\author{G.~Sangiovanni}
\affiliation{Max-Planck Institut f\"ur Festk\"orperforschung, Heisenbergstr. 1,
D-70569 Stuttgart, Germany}

\author{A.~Toschi}
\affiliation{Max-Planck Institut f\"ur Festk\"orperforschung, Heisenbergstr. 1,
D-70569 Stuttgart, Germany}

\author{E.~Koch}

\affiliation{Institut f\"ur Festk\"orperforschung, Forschungszentrum J\"ulich,
52425 J\"ulich, Germany}

\author{
K.~Held}
\affiliation{Max-Planck Institut f\"ur Festk\"orperforschung, Heisenbergstr. 1,
D-70569 Stuttgart, Germany}

\author{M.~Capone}

\affiliation{INFM-SMC and Istituto dei Sistemi Complessi, Consiglio Nazionale delle Ricerche, Via dei Taurini 19, I-00185
Roma, Italy}
\affiliation{
Dipartimento di Fisica Universit\`a di Roma "La Sapienza" piazzale Aldo Moro 5,
I-00185 Roma, Italy}

\author{C.~Castellani}

\affiliation{
Dipartimento di Fisica Universit\`a di Roma "La Sapienza" piazzale Aldo Moro 5,
I-00185 Roma, Italy}

\author{O.~Gunnarsson}
\affiliation{Max-Planck Institut f\"ur Festk\"orperforschung, Heisenbergstr. 1,
D-70569 Stuttgart, Germany}

\author{S.-K.~Mo}
\author{J.~W.~Allen}
\affiliation{Randall Laboratory of Physics, University of
Michigan, Ann Arbor, MI 48109}
\author{H.-D.~Kim}
\affiliation{Pohang Accelerator Laboratory, Pohang 790-784, Korea}
\author{A.~Sekiyama}
\author{A.~Yamasaki}
\author{S.~Suga}
\affiliation{Department of Material Physics, Graduate School of
Engineering Science, Osaka University, 1-3 Machikaneyama,
Toyonaka, Osaka 560-8531, Japan}
\author{P.~Metcalf}
\affiliation{Department of Physics, Purdue University, West
Lafayette, IN 47907}


\begin{abstract}
Studying the antiferromagnetic phase of the Hubbard model by
 dynamical mean  field theory, 
we observe striking differences with static (Hartree-Fock) mean field: 
The Slater band is strongly renormalized and
spectral weight is transferred to spin-polaron side bands.
Already for intermediate values of the interaction $U$ 
the overall bandwidth is larger than in Hartree-Fock, and the gap 
is considerably smaller.
Such differences survive any renormalization of $U$.
Our photoemission
experiments for Cr-doped V$_2$O$_3$ show
spectra qualitatively well described by  dynamical mean field theory.
 
\end{abstract}
\date{\today}
\maketitle

\section{Introduction}

In recent years,  our ability to  calculate strongly correlated materials has
substantially improved.
To this end, one needs to go beyond the conventional local density
approximation (LDA).\cite{LDA}
New methods had to be developed like  LDA+$U$,\cite{LDAU} where LDA is
supplemented by a local Coulomb interaction $U$ treated in the static
Hartree-Fock (HF) mean field theory, and its sibling LDA+DMFT,\cite{LDADMFT1,LDADMFT2} 
which employs the more sophisticated dynamical mean
field theory (DMFT).\cite{DMFT}
While it is generally accepted that LDA+DMFT deals more accurately with 
strongly correlated metals, the simpler LDA+U is considered to be sufficient 
for insulators with a large $U$,\cite{LDADMFT2} at least in the presence of
magnetic or orbital ordering.
Indeed, LDA+U was tailored for such strongly correlated insulators,\cite{LDAU} and 
is nowadays widely employed to calculate various physical quantities of these.

In our paper, we work out the differences between a static and a dynamical mean
field treatment of  long-range-ordered insulators at intermediate to strong
Coulomb interactions $U$.
For realistic material calculations, this would correspond to LDA+U and
LDA+DMFT.
We study these differences, considering the antiferromagnetic (AF) phase of a  
simple model, the one-band Hubbard model.
We point out that at large $U$ dynamic properties and
the structure of excited  states are strikingly different in the 
Slater (HF) and the DMFT antiferromagnet.
More specifically the Slater bands are strongly renormalized, 
most of the spectral 
weight is transferred to spin-polaron side bands, and the overall bandwidth is
proportional to the non-interacting width $W$, as opposed to the $1/U$ shrinking
found in HF.
We also performed photoemission spectroscopy (PES)
experiments for V$_2$O$_3$ doped with 1.2\% Cr, considering both the
AF and the
paramagnetic insulator. These experiments confirm
that strongly correlated antiferromagnets
are not of Slater (HF)  type,
while they can in many respects be described by DMFT.

The paper is organized as follows: In Sec. \ref{due} the DMFT results for the 
antiferromagnetic phase of the Hubbard model are presented and compared to 
the exact solution of the $t$-$J$ model in infinite dimensions. 
In Sec. \ref{tre} we analyze the evolution of staggered magnetization and 
spectral function from weak to strong coupling. This allows us to draw some general 
conclusions about the validity of DMFT and HF in describing both ground-
and excited-state properties in the antiferromagnetic phase. 
In Sec. \ref{quattro} the evolution from a Mott antiferromagnet to a 
paramagnetic Mott insulator with increasing temperature is discussed.
The photoemission spectrum of V$_2$O$_3$ is also presented in both phases 
and it is compared to our theoretical calculation. Finally we present our 
conclusions in Sec. \ref{cinque}.

\section{Antiferromagnetic phase of the Hubbard model} \label{due}

Several studies have been performed on the AF phase of 
the Hubbard model,\cite{AFDMFT} but here we focus on large values of $U$, 
that received less attention in the past, and we exploit some technical 
advances to improve the accuracy of our DMFT calculation 
 so that we can make definite statements not only about the size of the 
AF gap but also regarding the inner structure of the 
(Hubbard or Slater) bands below (and above)
the gap. In the large $U$ limit, we can make contact with the $t$-$J$ model
which has been studied in infinite dimensions by Strack and Vollhardt.\cite{volly} 
We recover the spin-polaron peaks of Ref. \onlinecite{volly} which are however 
dispersive in our calculation, an effect occurring in order $D^2/U^2$ and hence 
absent in Ref. \onlinecite{volly}.

The Hubbard Hamiltonian reads
\begin{equation} \label{hamiltonian}
H =  -t\sum_{\langle i,j\rangle, \sigma}  c^{\dagger}_{i \sigma} c_{j \sigma} +
H.c.
+ U \sum_i n_{i\uparrow} n_{i\downarrow}
\end{equation}
where $t$ is the hopping amplitude, $U$ is the Coulomb repulsion, $c_{i \sigma}$
($c^{\dagger}_{i \sigma}$) are annihilation (creation) operators for spin
$\sigma$ electrons on site $i$ and $n_{i \sigma}=c^{\dagger}_{i \sigma} c_{i
\sigma}$.
We solve the DMFT equations for a semicircular density of states
$N(\varepsilon)\!=\!(2/\pi D^2 )\;\sqrt{D^2-\varepsilon^2}$ with bandwidth $W=2D$,
using exact diagonalization and Lanczos algorithm at $T=0$ for the associated
Anderson impurity model (AIM).
A continued fraction expansion of the DMFT Weiss fields \cite{Kettenbruch}
allows us to obtain reliable spectra in the intermediate-to-large $U$ region. 
Moreover, in the AF phase we are able push the number of AIM sites $N_s$
to much larger values than for the paramagnetic phase.

\begin{figure}[tbp]
\begin{center}
\includegraphics[width=8.5cm]{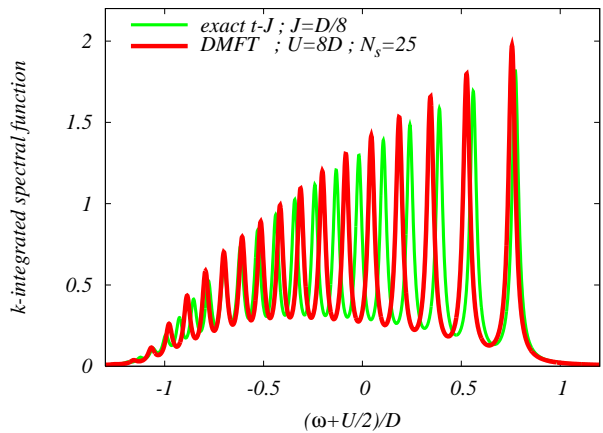}
\end{center}
\caption{(Color online) Spectral function of the lower Hubbard band. 
Red (dark gray) curve: DMFT solution of Hubbard model for $U=8D$ and $N_s=25$. 
Green (light gray): $t$-$J$ model for $J=D^2/U$. 
Both spectra have been plotted using the same Lorentzian broadening.} 
\label{fig1}
\end{figure}
To confirm that we get reliable spectral functions we first consider the
large-$U$ limit of the Hubbard model and compare it to the exact solution
for the $t$-$J$ model in infinite dimensions.\cite{volly}
The comparison of the $k$-integrated spectral functions is displayed in 
Fig.\ \ref{fig1}.
In both cases, we show only the negative
frequency part of the spectrum, i.e., the lower Hubbard band centered around
$\omega=-U/2$ (the $x$ axis is shifted correspondingly).
Evidently in both cases we find spectra of the same type. We have also
verified that the DMFT spectra are almost independent on $N_s$, reflecting the fact 
that, contrary to the case of the paramagnetic insulating phase, in the AF
phase the peaks do not originate from the discreteness of the impurity model
but have a physical meaning.\cite{BrinkmanRice,volly}

As already observed in Ref. \onlinecite{volly} they originate from the fact that 
a hole moving in the ordered background breaks AF bonds costing an
energy proportional to $nJ$, where $n$ is the length
of path in the lattice. 
This string potential gives rise to a set of discrete energy levels
with a typical separation proportional to $J^{2/3}$. 
These levels can be interpreted as spin-polaron side peaks\cite{spinpolaron} 
for dispersionless spin-waves.\cite{cappciuk}
In the $U\!\rightarrow\!\infty$ ($J\!\rightarrow\!0$) limit, these spin-polaron
peaks become dense and the  lower Hubbard band recovers the shape
and the width of the non-interacting density of states, with the
important difference that the states are incoherent 
for  $U\!\rightarrow\!\infty$. 
In our finite-$U$ case, we find that the shift of the peaks
in the Hubbard model with respect to the $t$-$J$ model scales with $1/U^2$.
That means we recover the infinite-$U$ limit where the mapping of the
Hubbard to the $t$-$J$ model is exact.

\section{static and dynamical properties from weak to strong-coupling} \label{tre}

After having demonstrated the accuracy of our spectra we investigate how the 
picture changes when $U$ is lower and follow the evolution of our 
antiferromagnetic solution as a function of $U$.

\begin{figure}[tbp]
\begin{center}
\includegraphics[width=8.5cm]{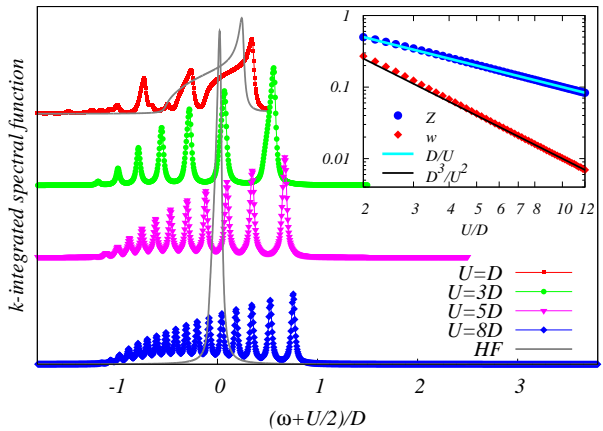}
\end{center}
\caption{(Color online) Evolution of the lower Hubbard band with $U/D$. 
The solid gray lines
represent the HF spectrum for $U=D$ and $U=8D$.
In the inset the spectral weight and the quasiparticle dispersion are
shown as a function of $U/D$}
\label{fig2}
\end{figure}

While, for $U = D$ static and dynamical mean-field theories yield, as one expects, 
similar spectra (see uppermost curve in the main panel of Fig. \ref{fig2}), 
already for $U=3D$ the spin-polaron picture characteristic of the $t$-$J$ model
is almost fully developed. This is instead completely missed by the static 
mean-field theory.
The discrepancy between DMFT and HF spectrum grows with increasing $U$, as 
exemplified in the figure for $U = 8D$. 
The HF spectrum arises from a renormalized single-particle band with 
a dispersion  decreasing as $1/U$. In contrast in DMFT the width of 
the Hubbard band is always of the order of the bare bandwidth even in the limit 
$U \to \infty$.

Focusing on the rightmost peak of the Hubbard band, we find that it 
continuously evolves from the weak-coupling Slater (HF) peak.
However, its spectral weight $Z$, directly measured from the spectral density,
 is dramatically reduced when increasing $U$ since more and more
weight is transferred to the spin-polaron side bands.
As it can be seen in the inset of Fig. \ref{fig2}, $Z$ scales as $D/U$ for 
$U \gtrsim 2D$. 
This is what is also found for the $t$-$J$ model, however the physical nature 
of the excitation is different. 
In the $t$-$J$ model the peaks are dispersionless,\cite{volly} while in the
Hubbard model we find, for the peak closest to the Fermi level, a bandwidth $w$ 
which scales as $D^3/U^2$ for large $U$ (solid diamonds in the inset of 
Fig. \ref{fig2}).

An independent argument for the dispersion of this peak is obtained by explicitly
looking for the poles of the Green's function.\cite{AFDMFT}
Averaging over the two sublattices we obtain
\begin{eqnarray} \label{green}
G(\varepsilon_k,\omega) & =& 
\frac{1}{2} \left(G_{AA}(\varepsilon_k,\omega) + G_{BB}(\varepsilon_k,\omega) \right) 
\nonumber \\
& =& \frac{\frac{1}{2}(\zeta_A + \zeta_B)}{\zeta_A \zeta_B - \varepsilon_k^2}
\end{eqnarray}
where $\varepsilon_k$ is the bare band energy,
$\zeta_{\alpha}=\omega + \mu - \Sigma_{\alpha}(\omega)$ and $\alpha=A$, $B$ is 
a sublattice index.
The renormalized dispersion $E_k$ is given by the solutions of 
\begin{equation} \label{poles}
\mbox{Re} \zeta_A(\omega) \zeta_B(\omega) - \varepsilon_k^2 = 0
\end{equation}
with $\varepsilon_k$ ranging from $-D$ to $D$.

We label $\omega_{qp}$ the solution for $\varepsilon_k=0$.  
At large $U$ we expand Eq.~(\ref{poles}) to linear order and obtain
\begin{equation}\label{EpskDMFT1}
E_k-\omega_{qp}={1\over \varkappa}\varepsilon_k^2
\end{equation}
where we have defined 
\begin{equation}\label{varkappa}
\varkappa \equiv  
\left. {\partial \over \partial \omega} \mbox{Re} \zeta_A(\omega)\zeta_B(\omega)
\right|_{\omega=\omega_{qp}}.
\end{equation}
To determine the value of $\varkappa$, we consider the Green's function
for energies $\omega \sim -U/2$, which for large $U$ is given by 
the Green's function of a given sublattice, e.g. $G_{AA}(\omega)$. 
In this limit, $\Sigma_{B}\approx -\Sigma_{A} \approx U/2$. 
Then we find that the quasiparticle weight is given by $Z=-U/\varkappa$.

In particular, considering as an example the case of a hypercubic lattice, 
the dispersion $E_k$ has the following form:
\begin{equation}\label{EpskDMFT2}
E_k-\mbox{const.}=-{Z\over U}\varepsilon_k^2=-{Z\over U}
(2t\sum_n {\rm cos} k_n)^2,
\end{equation}
where $k_n$ is the wave vector in the $n$th dimension.

The presence of a finite quasiparticle dispersion in the Hubbard model can be 
understood as an effect of the $\mathcal{O}(1/U^2)$ terms neglected in the 
mapping to the $t$-$J$ Hamiltonian.\cite{auerbach}
More precisely, in addition to the standard $t$-$J$ terms,
the large-$U$ projection of the Hubbard model leads to 
spin-flip terms and to second ($t_2$) and third ($t_3$) nearest neighbor 
hopping integrals. Such high-order hoppings contribute to the
quasiparticle dispersion even in infinite dimensions, and in the
specific case of a hole in a N\'eel state read $t_2= 2t^2/U$ and $t_3= t^2/U$,
respectively.
For a hypercubic lattice these terms give rise to a bare dispersion of the form 
\begin{eqnarray}\label{epskt2t3}
\tilde{\varepsilon}_k & = & 
-2t_2\sum_{n\ne m} \cos k_n \cos k_m -2t_3\sum_n \cos 2k_n  \nonumber \\
& = & -\frac{1}{U}( 2t\sum_{n}\cos k_n ) ^2 -  \mbox{const.}
\end{eqnarray}
From the results obtained in Ref. \onlinecite{vollytpri} for a $t$-$t'$-$J$
model, we expect for the Hubbard model, in the limit of $t_2\ll 2J$ and $t_3\ll 2J$, 
a dispersion given by the bare $\tilde{\varepsilon}_k$ of Eq. (\ref{epskt2t3}) 
renormalized by the quasiparticle weight $Z\simeq D/U$.
This expectation is in fact confirmed since $Z\tilde{\varepsilon}_k$ 
coincides with the dispersion given by Eq. (\ref{EpskDMFT2}).

These considerations support the interpretation of our DMFT results as a 
quasiparticle renormalization of the Slater (HF) peak; the width of this peak 
shrinks like $D^2/U$ and it is further reduced by $Z\simeq D/U$ so that 
altogether $w\!\simeq \!Z D^2/U \!=\!D^3/U^2$.

\begin{figure}[tbp]
\begin{center}
\includegraphics[width=8.5cm]{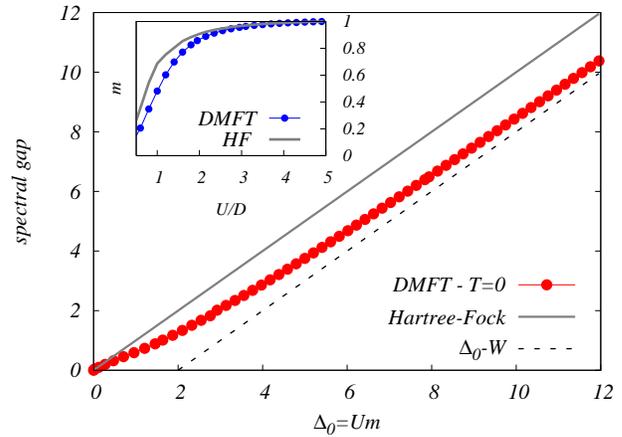}
\end{center}
\caption{(Color online) DMFT spectral gap as a function of the AF order parameter 
$\Delta_0=U m$, where $m$ is the staggered magnetization. 
In the Hartree-Fock case the gap coincides with $\Delta_0$ (dashed line). 
In the inset we compare the DMFT and the HF magnetization as a function of $U$.}
\label{newfig}
\end{figure}

As already mentioned before, a striking difference between the Slater and the DMFT spectrum is the
amplitude of the spectral gap. Since the HF Slater 
peak is located roughly at the center  of the DMFT lower Hubbard band of width of order $W$,
for large $U$ the spectral gap is reduced from the HF value $U$ to $U-W$.
We illustrate this aspect in Fig. \ref{newfig}, where the amplitude of the spectral
gap is plotted as a function of the order parameter $\Delta_0$. $\Delta_0$ is given
by $U$ multiplied by the staggered magnetization $m$, whose behavior is shown in 
the inset of Fig. \ref{newfig}. 
The DMFT results indicate that, on the one hand the staggered magnetization gets 
closer and closer to the Hartree-Fock value by increasing $U$, while, on the
other hand, the spectral gap deviates more and more from the Hartree-Fock 
prediction ($\Delta_0$) approaching the asymptotic value of $\Delta_0-W$. 
Thus we have the interesting situation where the two methods give essentially the
same magnetization, while there is a sizeable difference in the gaps.

Our findings highlight one important point:
At large $U$, DMFT and HF give the same description of the antiferromagnetic 
ground-state, which is basically the N\'eel state. 
On the other hand, the description of excited states is strikingly different 
between DMFT and HF.
In HF the hole moves in a ``rigid'' background of N\'eel spins, while, in DMFT, due 
to quantum fluctuations, the excitations are of a completely different nature: The 
hole can also move around without paying the double occupancy cost, with a creation 
of a spin polaron. 
Therefore, even if all ground-state quantities ($m$, kinetic and potential energy, 
\ldots) are basically the same in the two descriptions, excited state properties 
(hence spectra) are completely misrepresented by HF. 

This also means that, within HF, it is not possible to simultaneously obtain the 
correct magnetization and spectral gap with the same value of $U$. 
This has obvious implications for the use of LDA+U for  antiferromagnets
(or similarly for orbitally-ordered phases) with a large gap.

\section{antiferromagnetic vs paramagnetic Mott insulator} \label{quattro}

\begin{figure}[tbp]
\begin{center}
\includegraphics[width=8.5cm]{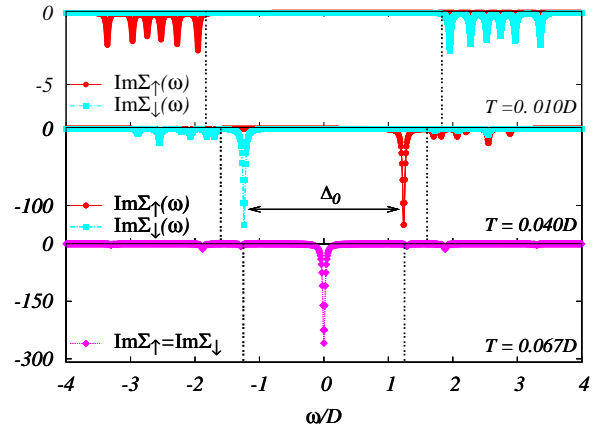}
\end{center}
\caption{(Color online) Imaginary part of up and down self-energy for $U=5D$, below (two upper
panels) and above (lowest panel) the N\'eel temperature ($T_N=0.05D$). The
spectral gap is marked by the dashed vertical lines and $\Delta_0=Um$.}
\label{fig3}
\end{figure}

Our DMFT calculations can be extended also to finite temperature. In this case,
calculations can only be performed with a lower value of $N_s$, but their
precision is still sufficient to address a crucial issue of the Mott
antiferromagnet, namely the evolution from a $t$-$J$ antiferromagnet to a 
paramagnetic Mott insulator above the N\'eel temperature.
In particular we are interested in the mechanism ruling the evolution of the
system from a fully polarized antiferromagnet with a spectral gap of order $U$ to a Mott
insulator with zero magnetization but with 
a gap of the same order of magnitude.
At low $T$, due to the onset of long-range AF order, the opening of the gap is
accompanied by a rigid shift in the real parts of the self-energy, 
whose value is given by the magnitude of the order parameter.
The imaginary parts are, in this case, small for both spin species and non-zero
only where the corresponding (up or down) spectral weight is present (upper
panel of Fig.\ \ref{fig3}).
On the other hand, for $T > T_N$, when the order parameter is zero, a Mott gap of
order $U$ is found, associated with a $\sim U^2/\omega$ peak in the imaginary
parts of $\Sigma(\omega)$ at zero frequency, as can be seen in the lowest
panel of Fig.\ \ref{fig3}.
How can we go from one situation to the other?
By increasing $T$, two peaks in the imaginary parts of $\Sigma(\omega)$ develop
inside the spectral gap (indicated by the vertical dashed lines in the figure)
and their mutual distance decreases together with the order parameter
$\Delta_0$:
The two peaks can be viewed as precursor of the $\sim U^2/\omega$ 
behavior in the paramagnetic Mott insulator.

\begin{figure}[tbp]
\begin{center}
\includegraphics[width=9.0cm]{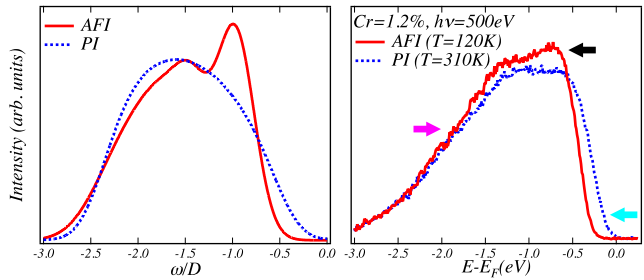}
\end{center}
\caption{
(Color online) Comparison of theory (left) and PES (right) spectra of the lower 
Hubbard band in AF and paramagnetic phases. 
The black, cyan (light gray), and magenta (dark gray) arrows, indicate, respectively, 
the formation of the (quasiparticle-)renormalized Slater peak, the reduction of the 
gap in the AF phase and the qualitative disagreement in the lower edge of the Hubbard 
band.}
\label{fig4}
\end{figure}

Let us now ask: Are real transition metal oxides described by static or dynamical mean field  theory?
To this end, we performed photoemission measurements for V$_2$O$_3$ doped with 
1.2\% Cr. \cite{JimAFIPI} The measurement was made at beamline BL25SU of SPring-8 using 
500eV photons. The experimental details are exactly as described previously.\cite{Mo}
The comparison between theory and experiments is shown in Fig. \ref{fig4}, taking parameters 
reasonable for V$_2$O$_3$: $U=3$[eV] and $2D=2$[eV]. 
As we discuss below, the theoretical spectra have been broadened to make contact with 
the experiment. Both theoretical and experimental spectra are normalized to have the same area. 
A Shirley-type inelastic background has been removed from the experimental data. 

Comparing the experimental data (right panel of Fig.(\ref{fig4})) to the spectra we have 
discussed until now, it is evident that the spiky nature of the latter is absent, or at 
least basically invisible, in the former. 
This is evidently due to at least three sources of broadening: the intrinsic experimental
broadening, the finite temperature and finite dimensionality effects beyond the DMFT.
This last source of broadening can, e.g., arise from the coupling to 
{\em dispersive} spin-waves.\cite{spinpolaron} 
The size of this effect is hard to estimate for three-dimensional compounds.
However, to mimic the overall effect we have plotted in the left panel of 
Fig.(\ref{fig4}) our theoretical results using a broadening of $0.2$[eV], larger 
than the experimental one ($\sim 0.1$[eV]) in order to include phenomenologically 
the other effects. 

A first observation is that the experimental spectrum in the antiferromagnetic phase
has an overall width of the same order of the paramagnet. We have already shown
how this is realized in the DMFT spectra, in contrast to the static HF mean-field.
Theory and experiment qualitatively agree also on the substantial redistribution of 
spectral weight: Going from the paramagnetic to the AF insulator, an additional peak 
develops at the upper edge of the lower Hubbard band. This peak is the one we identified 
as a (quasiparticle-) renormalized Slater band. Besides this peak there is additional 
spectral weight coming from the broadened spin polaron side-bands.
Remarkably, the gap is found to increase when going from the paramagnetic to the AF
insulator both in experimental and theoretical spectra. 
It is worth remarking that this feature does not depend on a particular choice of
the value of $U$, since it occurs even in the infinite-$U$ case.\cite{MetVoll}
The agreement is less good in one qualitative aspect: 
the flank at the lower edge of the Hubbard band is
almost identical for both phases in experiment whereas they
slightly differ in the theoretical calculation. 
This is likely due to the lack of some realistic features, like an asymmetric 
density of states and multi-orbital effects in  our model calculations.
This point could be clarified by extending LDA+DMFT calculations 
from the paramagnetic phase \cite{Held01} to AF V$_2$O$_3$. 

\section{conclusions} \label{cinque}

In this work we have compared static and dynamical mean field descriptions
of the antiferromagnetic phase of the Hubbard model. We have analyzed the evolution
from weak to strong coupling of both ground-state (e.g. magnetization) and 
excited-states (e.g. spectra) properties. Contrary to the general expectation
that HF gives a good description of broken-symmetry phases both at weak and 
strong coupling, we have found that, at large $U$, this is true only for 
ground-state quantities.
The poor description of the excitations given by static-mean field reflects in a
HF spectrum given only by two unrenormalized ($Z=1$) Slater bands which shrink as 
$1/U$ with increasing $U$. 
On the contrary, when the excited states are better described, as in DMFT, the 
Slater bands get strongly renormalized ($Z \simeq D/U$) and most of the spectral 
weight is transferred to spin polaron side bands. 
Therefore the total width of the Hubbard band stays finite (of order $W$) even for 
large $U$ and the spectral gap is not given by $U$, as in HF, rather by $U-W$.

We have also performed PES experiments on Cr-doped V$_2$O$_3$ both in the 
antiferromagnetic and paramagnetic phase. We have found a qualitative agreement with 
the DMFT calculations, in the formation of the renormalized Slater peak, in the
overall width of the band (of order $W$ rather than $D/U$), and also in the shrinking 
of the gap when going from the paramagnetic to the antiferromagnetic phase. 

In conclusion, we demonstrated with our DMFT calculation and by comparing theory with 
experiments that one has to be careful when applying static mean field theories, 
like LDA+$U$, for calculating 
physical properties of insulators with long range order.
Adjusting $U$ to get the correct gap, one obtains wrong estimates of the 
magnetization, or the energy, and {\em vice versa}. 
These problems can be overcome by employing dynamical mean field theory.

\section{acknowledgments}

This work was supported by the
Emmy-Noether program of the Deutsche
Forschungsgemeinschaft (AT,KH), by U.S. NSF (Grant No.\ DMR-03-02825), by a Grant-in-Aid for COE Research (10CEW2004)
and by MIUR PRIN Cofin 2003.
GS and MC acknowledge hospitality of Forschungszentrum J\"ulich. 
Calculations were performed on the J\"ulich JUMP computer under Grant No.\ JIFF22


\end{document}